\documentclass[final] {aipproc}
\usepackage{sidecap,graphicx,boxedminipage,epsfig,wrapfig,floatflt,shadow}
\layoutstyle{8x11double}

\begin{document}

\title{Students' Understanding of Stern Gerlach Experiment}

\author{Guangtian Zhu and Chandralekha Singh}{
  address={Department of Physics and Astronomy, University of Pittsburgh, Pittsburgh, PA, 15260, USA}
}

\begin{abstract}
The Stern Gerlach experiment has played a central role in the discovery of spin angular momentum and it 
has also played a pivotal role in elucidating foundational issues in quantum mechanics.
Here, we discuss investigation of students' difficulties related to the Stern Gerlach experiment
by giving written tests and interviewing advanced undergraduate and graduate students in quantum mechanics.
We also discuss preliminary data that suggest that the Quantum Interactive Learning Tutorial (QuILT) related to the Stern
Gerlach experiment is helpful in improving students' understanding of these concepts.
\end{abstract}

\maketitle

\section{Introduction}
\vspace*{-.08in}

Learning quantum mechanics is challenging~\cite{zollman,wittman,my2,sam2}. We are developing and evaluating Quantum Interactive Learning Tutorials (QuILTs)
to help students in advanced undergraduate quantum mechanics courses~\cite{singh}. 
QuILTs are based upon research on students' difficulties in learning quantum mechanics.
They strive to build on students' prior knowledge, actively engage them in the learning process and help them build links between 
the abstract formalism and conceptual aspects of quantum physics without compromising the technical content.

Here, we discuss investigation of students' difficulties related to the Stern Gerlach experiment (SGE) and the development and preliminary
evaluation of a QuILT that focuses on helping students learn about foundational issues in quantum mechanics via the Stern Gerlach experiment.
In the SGE, a particle with spin and/or orbital angular momentum (i.e., particle with a magnetic dipole moment) is sent through a 
Stern Gerlach Apparatus (SGA) with a non-uniform magnetic field. With appropriate gradient of the external magnetic field, different components of 
the angular momentum in the
wave function can be spatially separated by coupling them with different linear momenta. By using suitable measurement devices (e.g., detectors
at appropriate locations in the path of the beam), we can use the SGE to prepare a quantum state which is different from the initial state
before the particle entered the SGA. 

The investigation of difficulties was carried out by administering written surveys to more than two hundred physics graduate 
students and advanced undergraduate students enrolled in quantum mechanics courses and
by conducting individual interviews with a subset of students.
The individual interviews were carried out using a think aloud protocol
to better understand the rationale for their responses before and after the preliminary version of the QuILTs and the
corresponding pre-test and post-test were
developed. During the semi-structured interviews, students were asked to verbalize their thought processes while they answered
questions either as a part of the QuILT or as separate questions before the preliminary version of the QuILT was developed. 
Students were not interrupted unless they remained quiet for a while. 
In the end, we asked them for clarifications of the issues they had not made clear earlier. 
After each individual interview with the preliminary version of the QuILT (along with the pre-test and post-test), 
it was modified based upon the feedback. Our approach to the development of QuILTs is similar to the development of the tutorials
by the University of Washington Physics Education group. When we found that the QuILT on SGE was working well in individual administration and post-test performance
was significantly improved compared to pre-test performance,
it was administered in an advanced quantum mechanics class.
The QuILT comes with a warm-up exercise and
homework that students work on before and after working on the QuILT in class, respectively.

\vspace*{-.25in}
\subsection{Student Difficulties with SGE}
\vspace*{-.09in}

One question we have asked many first year physics graduate students and advanced undergraduate students related to the SGE 
in written tests and interviews is the following: \\
{\it $\vert \uparrow \rangle_z$ and  $\vert \downarrow \rangle_z$ represent the 
orthonormal eigenstates of $\hat S_z$ ($z$ component of spin angular momentum). SGA is an abbreviation for a
Stern-Gerlach apparatus.}

\begin{figure}
\epsfig{file=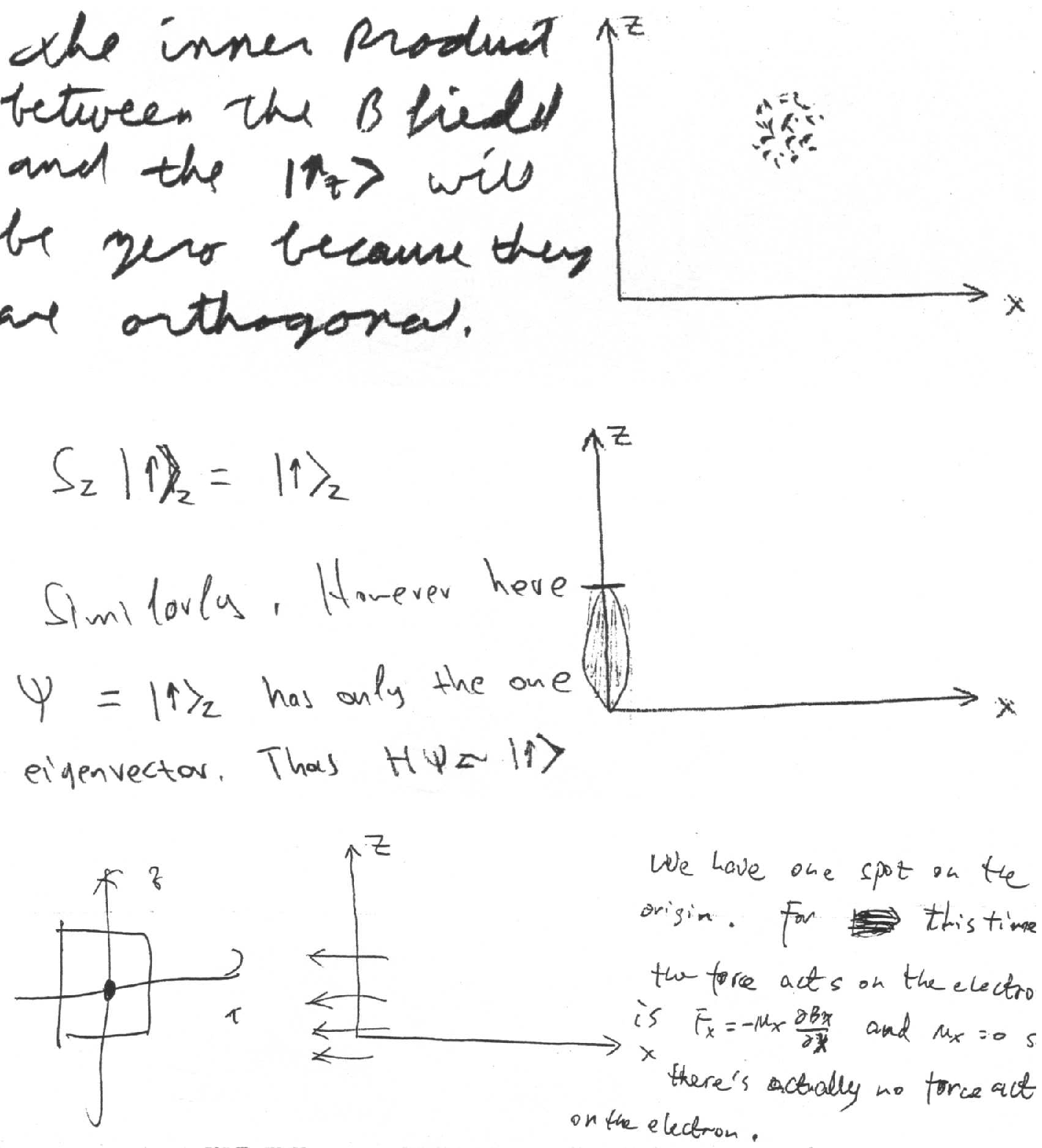,height=3.46in}
\caption{Three sample responses in which students provided incorrect explanations for why there should be one spot instead of two.} 
\end{figure}
\noindent
{\it \underline{\bf Question:} A beam of neutral silver atoms propagating along the $y$ direction (into the page) in spin state 
$\vert \uparrow \rangle_z$
is sent through a SGA with a horizontal magnetic field gradient in the $-x$ direction.
Sketch the pattern that you expect to observe on a distant phosphor screen in the x-z plane when the atoms hit the screen. Explain your reasoning.}

This question is challenging because students have to realize that, since the magnetic field gradient is in the $-x$ direction, the basis
must be chosen to be the eigenstates of $\hat S_x$ to readily analyze how the SGA will affect the spin state.
Here, the initial state, which is an
eigenstate of $\hat S_z$, $\vert \uparrow \rangle_z$, can be written as a linear superposition of the eigenstates of $\hat S_x$, i.e.,
$\vert \uparrow \rangle_z=(\vert \uparrow \rangle_x + \vert \downarrow \rangle_x) /\sqrt{2}$.
The magnetic field gradient in $-x$ direction will couple the $\vert \uparrow \rangle_x$ and $\vert \downarrow \rangle_x$ 
components in the incoming spin state $\vert \uparrow \rangle_z$
with oppositely directed x-component of momentum and will cause two spots on the phosphor screen separated along the x axis.

Many students claimed that there cannot be any splitting of the beam
since the magnetic field gradient is in the $-x$ direction but the spin state is along the $z$ direction and they are orthogonal to each other
(see Figure 1).
Students' written responses and individual interviews with a subset suggest that they were incorrectly connecting the direction of the magnetic field in 
the physical space with the ``direction" of state vectors in the Hilbert space.
While many students thought that there will be only one spot on the screen, there was no consensus on the direction of deflection.
Some students drew the spot at the origin; some showed deflections along the positive or negative x direction; some drew deflections along the 
positive or negative z direction.
Some students incorrectly believed that the spin state in this situation will get pulled in one direction because the magnetic field gradient
is in a certain direction (see Figure 2). Asking the interviewed students explicitly about whether they could consider a basis that may be particularly 
appropriate for analyzing this problem was not helpful.

One interviewed student drew a diagram of a molecular orbital with several lobes (see Figure 3) and wondered what the 
plot of $\vert \uparrow \rangle_z$ would look like.
Another student drew the diagram shown in Figure 4 and described Larmor precession of spin but
did not mention anything about the force on the particle due to the non-uniform magnetic field as in the SGE. 
Written responses and
interviews suggest that many students were unclear about the fact that in a uniform external magnetic field, the spin will only precess 
(if not in a stationary state) but in a non-uniform magnetic field as in the SGE, there will be a force that can spatially separate the 
components of angular momentum in the wave function under suitable condition. 
Moreover, similar to the difficulty
of these students, we have found that many students have difficulty realizing that spin is not an orbital degree of freedom.

\begin{figure}
\epsfig{file=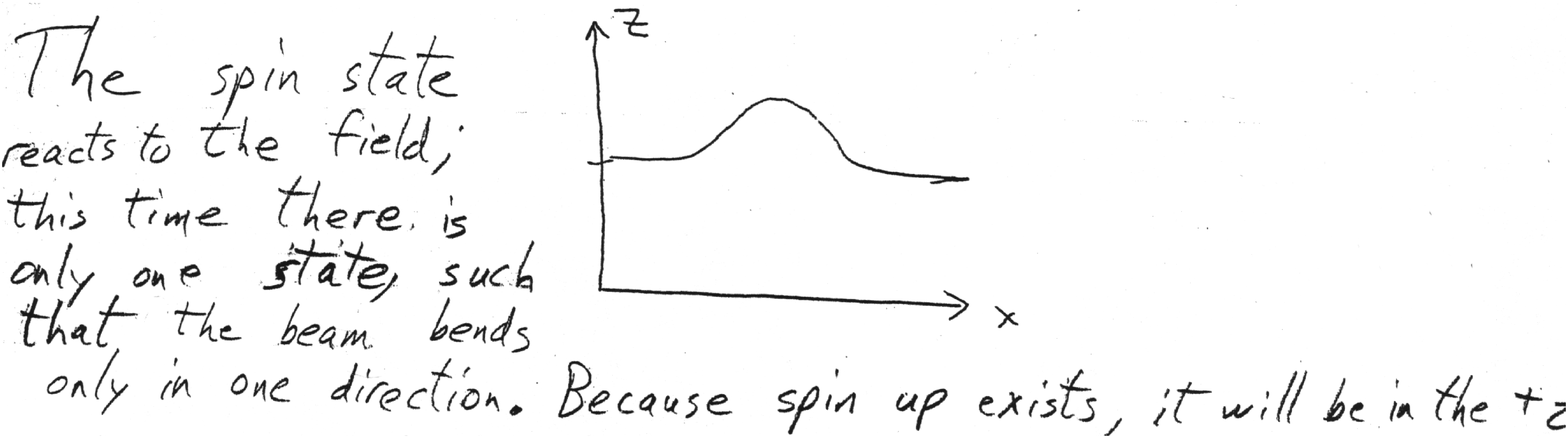,height=0.87in}
\end{figure}

\begin{figure}
\epsfig{file=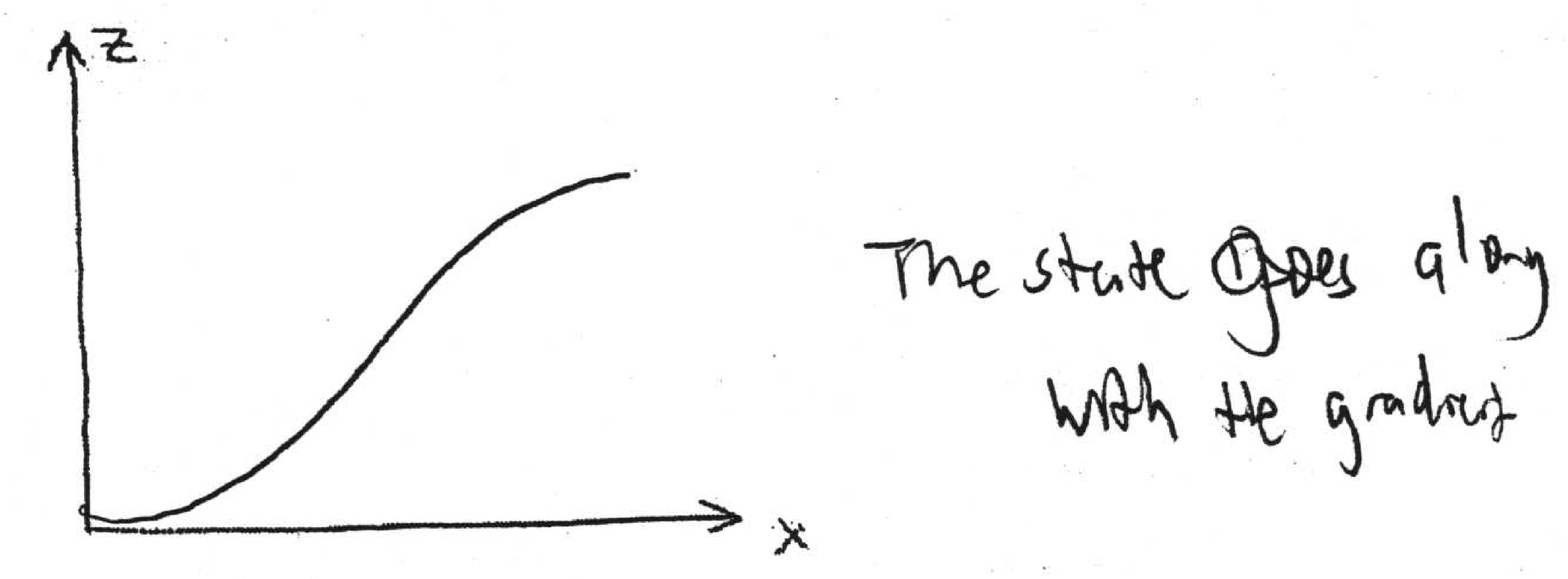,height=1.1in}
\caption{Two sample responses in which students provided incorrect explanations for why the state/beam will bend as shown in response
to the magnetic field gradient.} 
\end{figure}

\begin{figure}
\epsfig{file=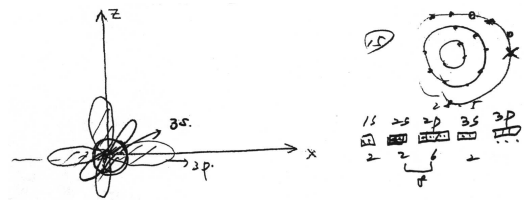,height=1.06in}
\caption{A student's diagram drawn to answer the question using the orbital picture}
\end{figure}

\begin{figure}
\epsfig{file=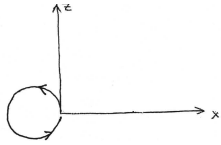,height=0.716in}
\caption{A diagram drawn by a student in response to the SGE question asked. In individual discussion, he described it as showing the 
Larmor precession of spin. 
The student incorrectly believed that spin is due to motion is real space.
Also, when he was reminded that the question was not about the dynamics (as suggested by the arrows drawn by the student
to show the direction of precession) but about the pattern observed on the screen,
he incorrectly claimed that the pattern would be a circle due to the precession.}
\end{figure}

In the interviews, students were also asked whether they could think of a strategy to distinguish between a superposition
in which particles are in $(\vert \uparrow \rangle_z + \vert \downarrow \rangle_z) /\sqrt{2}$ state
from a mixture in which half of the particles are in state $\vert \uparrow \rangle_z$ and the other half are in state $\vert \downarrow \rangle_z$.
This task was difficult for most students. One strategy for distinguishing between the superposition and mixture given
is to pass each of them one at a time through a $SGX_-$ (SGA with field gradient in negative x direction). Then, 
since state $(\vert \uparrow \rangle_z + \vert \downarrow \rangle_z) /\sqrt{2}$ is the same as $\vert \uparrow \rangle_x$, it will completely go out through the 
upper-channel after passing through $SGX_-$. On the other hand, the equal mixture of $\vert \uparrow \rangle_z$ and 
$\vert \downarrow \rangle_z$ will
have an equal probability of hitting detectors in both channels after the $SGX_-$ because these
states can be written as $(\vert \uparrow \rangle_x \pm \vert \downarrow \rangle_x) /\sqrt{2}$ in terms of the eigenstates of $\hat S_x$ and 
will become spatially separated after passing through the $SGX_-$.

\vspace*{-.22in}
\section{SGE QuILT}
\vspace*{-.09in}

As discussed in the introductory section, the QuILT about SGE builds on the prior knowledge of students investigated via written tests and interviews.
It uses a guided approach to help students build a knowledge structure about the SGE.
It uses computer-based visualization tools to help students build a physical intuition about relevant concepts related to the SGE. The
Open Source Physics SPINS program~\cite{mario} was adapted as needed throughout the SGE QuILT and can be easily tailored to the 
desired situations. Before each simulation, students predict what they expect and then reconcile the difference between their prediction
and observations. 
Toward the end, students are asked open-ended questions in which they have to
use the SPINS program to design experiments to accomplish certain tasks (some in more than one way).
The QuILT about SGE helps students learn about issues related to measurement, preparation of a desired quantum state, e.g.,
$\vert \uparrow \rangle_x$ starting from an arbitrary initial state, time-development of the wave function corresponding to the spin degree of freedom,
difference between superposition and mixture, difference between physical space and Hilbert space, importance of choosing an appropriate basis
to analyze what should happen in a given situation, etc. 
As an example, students are asked a sequence of questions such as the following to help them distinguish between superposition and mixture: 

{\it $\bullet$ If a silver atom in the state $( \vert \uparrow \rangle_z+\vert \downarrow \rangle_z)/\sqrt{2}$ (which is
an eigenstate of $\hat S_x$ with an eigenvalue $\hbar/2$) is passed through a $SGX_-$, what is the
probability that the ``up" detector will click? Explain.

$\bullet$ If an unpolarized mixture of silver atoms, half of which are in the state $\vert \uparrow \rangle_z$ and the
other half in $\vert \downarrow \rangle_z$, are passed through an $SGX_-$, write each component in the
mixture in a basis most suitable for analysis after passing through the $SGX_-$. Then write down the
probability that the ``up"  detector will click. Explain.}

Another question in the SGE QuILT asks students about preparing a $\vert \downarrow \rangle_z$ state
starting from a $\vert \uparrow \rangle_z$ state which is orthogonal to it.\\
{\it $\bullet$ You send silver atoms in an initial spin state $\vert \chi (0) \rangle = \vert \uparrow \rangle_z$ one at a time
through two SGAs with magnetic field gradients as shown below. Suitable detectors are
placed as shown. One detector is between the two SGAs and the other after both SGAs.
What is the probability that a given single atom will cause the ``up" detector to click after passing through this system of two SGAs?
If the detector does not click, what state have you prepared that you can collect in the lower channel?}
\begin{center}
\epsfig{file=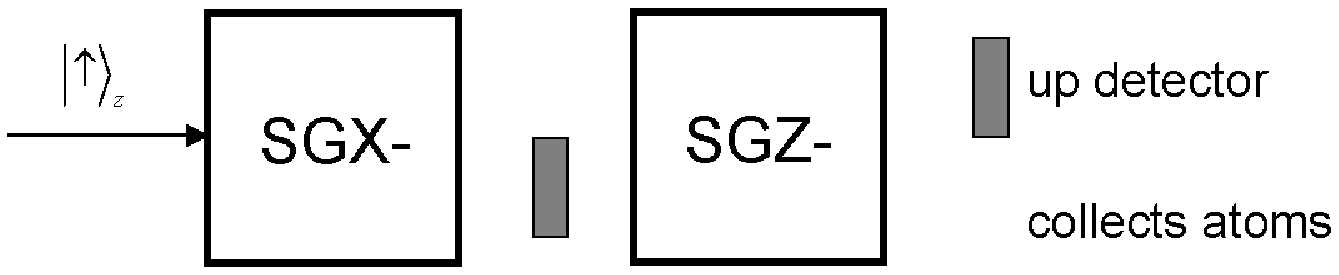,height=.7in}
\end{center}
\vspace*{-0.05in}
\begin{figure}
\epsfig{file=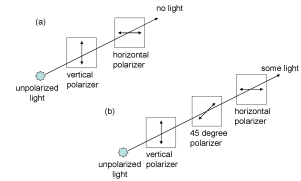,height=1.9in}
\caption{Analogy with photon polarization states.}
\end{figure}
\vspace*{-0.05in}
In order to help students understand that it is possible to input $\vert \uparrow \rangle_z$ through SGAs
and prepare an orthogonal state $\vert \downarrow \rangle_z$ on the way out, the QuILT also draws an analogy with photon polarization
states. Students learn that if you let atoms in the state $\vert \uparrow \rangle_z$ pass through SGZ only, you
will never obtain $\vert \downarrow \rangle_z$ on the way out. However, $\vert \downarrow \rangle_z$ is obtained
in the above experiment because we have inserted $SGX_-$ at an intermediate stage. Students consider the analogy with
vertically polarized light passing directly through a horizontal polarizer (Figure 5 a) vs.
passing first through a polarizer at $45^0$ followed by a horizontal polarizer (Figure 5 b).
There is no light at the output if vertically polarized light passes directly through a horizontal polarizer.
On the other hand, if the polarizer at $45^0$ is present, light becomes polarized at
$45^0$ after the $45^0$ polarizer, which is a linear superposition of horizontal and vertical polarization.
Therefore, some light comes out through the horizontal polarizer placed after the $45^0$ polarizer. 
Since the experiment with polarizers is familiar to students from introductory physics, this analogy helps students
learn about SGE using a familiar context.

We conducted preliminary evaluation of the QuILT in a class with 22 undergraduate students. Students had traditional instruction about
the SGE, took the pre-test, then worked on the tutorial and then took the post-test (some questions from the pre-test and post-test are in the
 Appendix). The average pre-test score was $53\%$ and the average post-test score was $92\%$, which is encouraging.

\vspace*{-.22in}
\section{Summary}
\vspace*{-.07in}

We have investigated students' difficulties related to SGE and used the findings as a guide to develop an SGE QuILT.
Preliminary evaluation suggests that the QuILT is helpful in improving students' understanding of concepts related to SGE.

\vspace*{-.23in}
\begin{theacknowledgments}
We thank NSF for 
awards PHY-0653129 and 055434.
\end{theacknowledgments}
\vspace*{-.15in}

\bibliographystyle{aipproc}

\vspace*{.2in}

{\bf Appendix: Examples of Pre-/Post-test Questions}

\vspace*{.1in}

\noindent
Pre-test Question (1): Chris sends silver atoms in an initial spin state $\vert \chi(0)  \rangle = (\vert \uparrow \rangle_z+ 
\vert \downarrow \rangle_z)/\sqrt{2}$
one at a time through an $SGX_-$. He places a ``down" detector in appropriate location as shown. What is the probability of the detector 
clicking when an atom exits the $SGX_-$?
\begin{center}
\epsfig{file=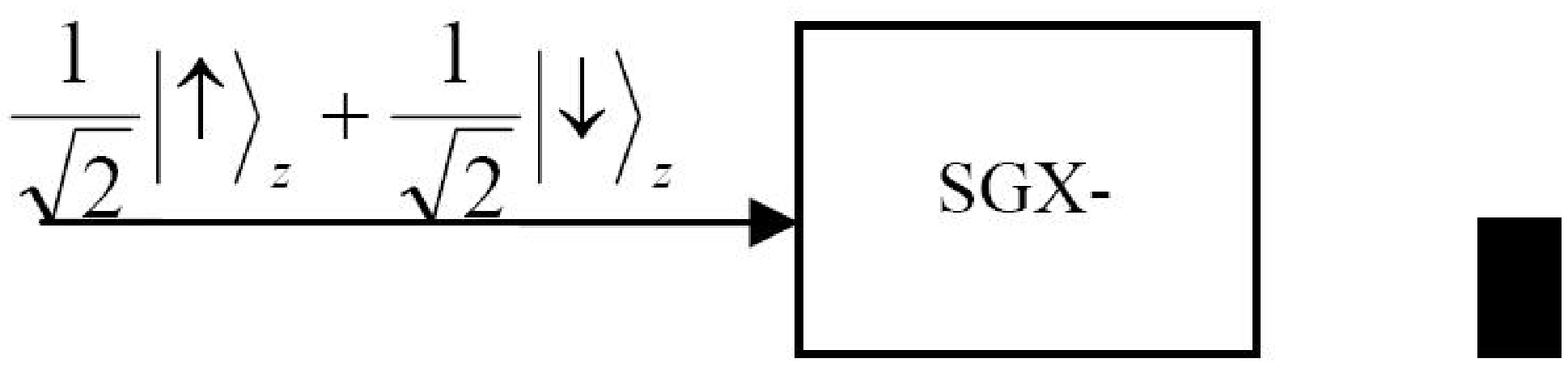,height=.76in}
\end{center}

\noindent
Pre-test Question (2): Silver atoms in an initial spin state $\vert \chi(0)  \rangle = \vert \uparrow \rangle_z$ pass one at a time
through two SGAs with the magnetic field gradients as shown below. Two suitable detectors are placed, one after the
first SGA and the second at the end to detect the atoms after they pass through both SGAs. The atoms that do not
register in the ``up" detector at the end are collected for another experiment. Find the fraction of atoms that
are detected in the ``up" detector at the end and the normalized spin state of the atoms that are collected for another experiment.
\vspace{-0.05in}
\begin{center}
\epsfig{file=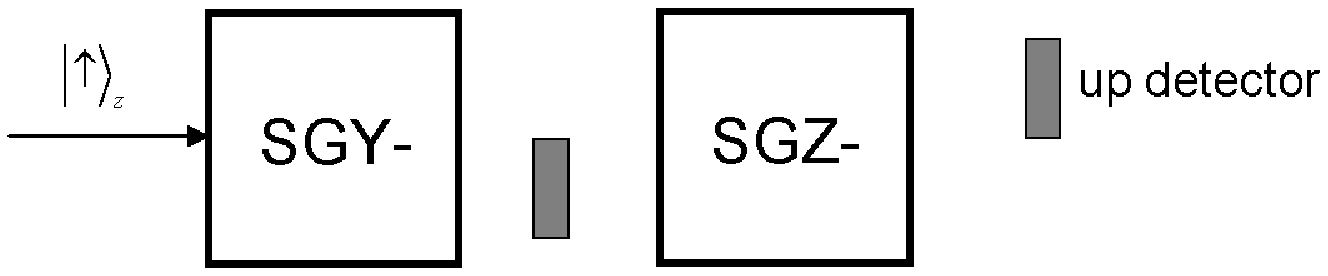,height=.7in}
\end{center}
\vspace{-0.05in}

\noindent
Post-test Question (1): Suppose beam A consists of silver atoms in the state 
$\vert \chi(0) \rangle = (\vert \uparrow \rangle_z+ \vert \downarrow \rangle_z)/\sqrt{2}$, and beam B is an unpolarized mixture in which half of 
the silver atoms are in state $\vert \uparrow \rangle_z$ and half are in state $\vert \downarrow \rangle_z$. 
Choose all of the following statements that are correct.\\
\noindent
\hspace*{0.45in} (1) Beam A will not separate after passing
\hspace*{0.45in} through SGZ (either $SGZ_-$ or $SGZ_+$).\\
\hspace*{0.45in} (2) Beam B will split into two parts after 
\hspace*{0.45in} passing through SGZ.\\
\hspace*{0.45in} (3) We can distinguish between beams A and 
\hspace*{0.45in} B by passing each of them through SGX.\\
A. (1) only\\
B. (2) only\\
C. (1) and (2) only\\
D. (2) and (3) only\\
E. All of the above

\noindent
Post-test Question (2): Sally sends silver atoms in state $\vert \uparrow \rangle_z$ through three SGAs as shown.
Next to each detector, write down the probability that the detector clicks. 
The probability for the clicking of a detector refers to the probability that a particle entering the {\bf first}
SGA reaches that detector.
Also, after each SGA, write the spin state sally has prepared. Explain.
\begin{center}
\epsfig{file=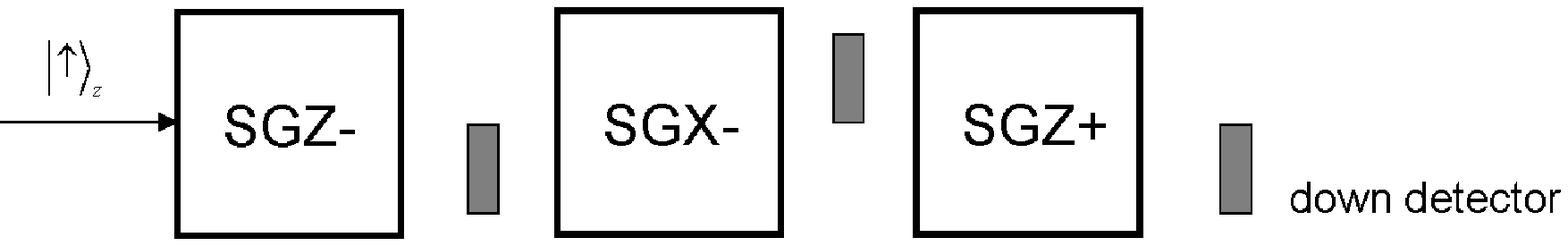,height=.5in}
\end{center}

\noindent
Post-test Question (3): Harry sends silver atoms all in the normalized spin state $\vert \chi \rangle =a \vert \uparrow \rangle_z +
b \vert \downarrow \rangle_z$ through an $SGX_-$.
He places an ``up" detector as shown to block some silver atoms and collects the atoms coming out in the ``lower channel"
for a second experiment. What fraction of the initial silver atoms will be available for his second experiment? What is the
spin state prepared for the second experiment? Show your work.
\begin{center}
\epsfig{file=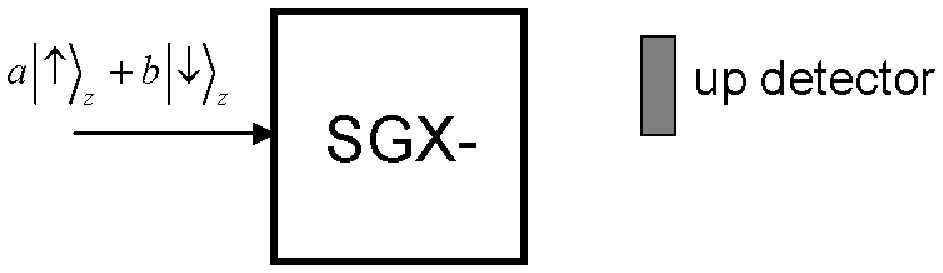,height=.7in}
\end{center}
\end{document}